\documentclass[pra,reprint,floatfix,showpacs,twocolumn, superscriptaddress]{revtex4-1}
\usepackage{amssymb,amsmath}
\usepackage{graphicx}
\usepackage{subfigure}
\usepackage{bm}
\usepackage[varg]{txfonts}

\def\be{\begin{equation}}
\def\ee{\end{equation}}
\def\bea{\begin{eqnarray}}
\def\eea{\end{eqnarray}}

\def\F1k{\widehat F_1({\bm k}^\prime)}

\def\tk1k{\widehat T_{{\bm k}_1,{\bm k}^\prime}}

\def\g1k{\widehat G^{-1}_{\bm k}}

\def\kk1{{{\bm k},{\bm k}_1}}

\def\bmatrix{\begin{pmatrix}}
\def\ematrix{\end{pmatrix}}

\begin{document}

\title{
Thermodynamic signatures for topological phase transitions to Majorana and Weyl superfluids in ultracold Fermi gases
}

\author{Kangjun Seo}
\affiliation{Department of Physics and Astronomy, Clemson University, Clemson, SC 29634, USA}
\author{Chuanwei Zhang}
\affiliation{Department of Physics, the University of Texas at Dallas, Richardson, TX 75080 USA}
\author{Sumanta Tewari}
\affiliation{Department of Physics and Astronomy, Clemson University, Clemson, SC 29634, USA}

\begin{abstract}
We discuss the thermodynamic signatures for the topological phase
transitions into Majorana and Weyl superfluid phases
in ultracold Fermi gases in two and three dimensions in the presence of
Rashba spin-orbit coupling and a Zeeman field. We analyze the thermodynamic
properties exhibiting the distinct nature of the topological phase
transitions linked with the Majorana fermions (2D Fermi gas) and Weyl
fermions (3D Fermi gas) which can be observed experimentally, including
pressure, chemical potential, isothermal compressibility, entropy, and
specific heat, as a function of the interaction and the Zeeman field at both
zero and finite temperatures. We conclude that among the various
thermodynamic quantities, the isothermal compressibility and the chemical potential as a
function of the artificial Zeeman field have the strongest signatures of the topological transitions in both two and three dimensions. 
\end{abstract}

\maketitle






\section{Introduction}

Ultracold fermionic quantum gases have many advantages for investigations of
quantum phenomena in a highly tunable way. The experimental ability to
control the interaction between the Fermi atoms via Feshbach resonances~\cite%
{chin-2010} enables to study the crossover phenomena \cite{giorgini-2008}
from the BCS superfluid of weakly bound Cooper pairs to the BEC of tightly
bound molecules~\cite{regal-2004}, universal properties in the unitary
regime~\cite{Stewart-2010}, vortices~\cite{Zwierlein-2005}, and so forth. In
addition, the experimental control of the internal spin states via
radio-frequencies provides the study of the emergence of new superfluid
phases in a Fermi gas with population imbalance which serves as an effective
Zeeman field~\cite{Zwierlein-2006,Partridge-2006,Liao-2010}. Another
new development enables to trap two dimensional fermions in the quantum
degenerate regime~\cite{Martiyanov-2010,Feld-2011}.

Furthermore, in recent experimental breakthroughs the spin-orbit coupling
(SOC) has been developed using two-photon Raman processes in bosonic as well
as fermionic gases~\cite%
{spielman-2011,Pan-2012,Engels-2013,Wang-2012,Zwierlein-2012}, which has
given rise to theoretical investigations on the new quantum states of matter~%
\cite{Chapman-2011,Gong-2011,Gong-2012,kseo-2012}. Thus far, the experimental
realization of SOC is an equal combination of Rashba~\cite{Rashba-1984} and
Dresselhaus~\cite{Dresselhaus-1955} types of SOC. It has been shown
theoretically that in the presence of Rashba SOC and a perpendicular Zeeman
field, 2D cold fermion systems with an $s$-wave Feshbach resonance can enter
into a topological superfluid phase with zero energy Majorana fermion
excitations localized at the order parameter defects such as vortices \cite%
{Gong-2012, Zhang-2008,Sato-2009}. It has also been shown that in 3D the corresponding system is
capable of supporting the intriguing Weyl fermion excitations \cite%
{Gong-2011}. The realization of ultra-cold fermionic gases with Rashba SOC
and a Zeeman field thus offers the tantalizing possibility of experimental
engineering of all three kinds of fundamental relativistic fermions, namely,
the Dirac, Weyl, and Majorana fermions \cite{Pal-2010}. In addition to these powerful
experimental and theoretical developments, there have been more recently
added to the list. One is the development of the methods for high precision
measurements of the thermodynamic properties such as the compressibility,
the chemical potential, the entropy, and the heat capacity for uniform $s$%
-wave Fermi superfluids~\cite{Navon-2010,Ku-2012}. In recent developments,
Ref.~[\onlinecite{Ku-2012}] has completely determined the universal
thermodynamics of the interacting Fermi gas across the $s$-wave superfluid
transition and has captured the onset of superlfuidity in the
compressibility, chemical potential, entropy and the heat capacity. The
question we address in this paper is what are the corresponding
thermodynamic signatures for interacting ultra-cold fermions across the
\textit{topological} superfluid transitions to phases supporting Majorana
fermions (2D) and Weyl fermions (3D). Our main conclusion is that among the various 
thermodynamic quantities, the isothermal compressibility and the chemical potential as a 
function of the artificial Zeeman field have the strongest signatures of the topological transitions. 
The signatures for the transition from the non-topological $s$-wave superfluid to the Majorana
superfluid phase in 2D are stronger than the corresponding signatures in 3D where the transition is in the Weyl 
superfluid phase. The other thermodynamic quantities such as the pressure, entropy, and specific heat have only
weak signatures of the topological transitions (that is, the transition between a non-topological superfluid and a topological superfluid as a function of the Zeeman field), even though they are useful to capture the topological superfluid to the normal phase (pair potential $\Delta=0$) transitions in both two and three dimensions.

In condensed matter systems, SOC, which results from the coupling between
the spin degrees of freedom and the orbital motion of a quantum particle,
plays an important role in understanding interesting physics associated with
the topological quantum phenomena including the spin Hall effect~\cite%
{Nagaosa-2010,Xiao-2010}, topological insulators \cite{Hasan}, and
topological superconductors~\cite{Qi-2011}. These phases generically arise as a
result of topological quantum phase transitions (TQPT) in the appropriate
systems. TQPTs separate distinct phases of matter which have exactly the
same symmetries and thus are not associated with any spontaneous symmetry
breaking in the underlying Hamiltonian. Yet, the phases separated by TQPTs
have distinct topological properties which can often manifest themselves in
the properties of the order parameter defects. For instance, it has been
pointed out recently that spin-orbit coupled electron- or hole-doped
semiconductor thin films (2D) or nanowires (1D) with proximity induced $s$%
-wave superconductivity and a suitably directed Zeeman field ($\Gamma $)
supports novel non-Abelian topological states with the order parameter
defects supporting localized topological zero-energy excitations called
Majorana fermions~\cite{Sau-2010,Sau-2010-PRB,Roman,Oreg,Mao1,Mao2} above a critical
Zeeman field ($\Gamma >\Gamma _{c}$). These theoretical developments in the
condensed matter system followed earlier similar proposals in the context of
ultra-cold fermions in two dimensions in the presence of artificial SOC and
a Zeeman field and with $s$-wave pairing interactions \cite{Zhang-2008}.
While the two dimensional cold fermion systems support the Majorana fermions
in the order parameter defects (but are otherwise completely gapped in the
bulk), the corresponding three dimensional systems have topologically
protected gapless nodes in the bulk supporting the relativistic Weyl fermion
excitations. With Rashba-type of SOC and an increasing strength of the
Zeeman field, the three dimensional system supports a series of 3D
topological quantum phase transitions~\cite{Volovik-2003} from a
non-topological superfluid state with fully gapped fermionic excitations to
a topological superfluid state with four protected Fermi points (i.e., gap
nodes) and then to a second topological superfluid state with only two
protected Fermi points~\cite{Gong-2011}. Such 3D topological gapless
superfluids with Weyl fermions, which result from quasiparticle excitations
in a linearized Hamiltonian near the nodes, is different from the 2D fully
gapped topological superfluids with Majorana fermions~\cite%
{Gong-2012,Sau-2010}. Recently, similar Weyl fermion physics in an analogous
3D condensed matter system has also been discussed in the context of
ferromagnetic superconductors ~\cite{Sau-2012}. The results on the
thermodynamic signatures of the topological transitions to Majorana and Weyl
superfluid phases discussed in this paper apply to both cold fermion and
condensed matter systems although our focus will mostly be ultra-cold
fermions where the methods for high precision measurements of the
thermodynamic quantities have recently been successfully developed.



The paper is organized as follows. In Sec. II, we derive the mean-field
order parameter and atom-number equations for ultra-cold fermions with
Rashba SOC and a Zeeman field. We present the zero temperature phase diagram
and the thermodynamic quantities as a function of interaction for both 2D
and 3D Fermi superfluids from the self-consistent numerical calculations in
Sec. III. Section IV provides the phase diagram and the corresponding
thermodynamic properties at finite temperature. Section V consists of
discussion and conclusions.


\section{Mean-field Theory}

The system we consider is a uniform $s$-wave fermionic superfluid with the
atom density $n = N/V$ in the presence of Rashba-type spin-orbit coupling
(SOC) in the $xy$ plane and a Zeeman field along the $z$ direction. The
dynamics of the Fermi gas can be described by the Hamiltonian $H = H_{0}+H_%
\text{int}, $ where the single particle Hamiltonian
\begin{equation}
H_{0} = \sum_{\gamma \gamma ^{\prime }}\int d \mathbf{r} c_{ \mathbf{r}
\gamma }^{\dagger} \left[ \xi I -i\hbar \alpha \left( \partial_{y}\sigma
_{x}-\partial_{x}\sigma _{y}\right) + \Gamma \sigma_{z} \right]_{\gamma
\gamma^{\prime}} c_{ \mathbf{r} \gamma^{\prime}},
\end{equation}
where $\gamma =\uparrow ,\downarrow $ are the pseudo-spin of the atoms, $\xi
= -\hbar^2\nabla^2 / 2m - \mu $, $\mu $ is the chemical potential, $\alpha$
is the Rashba SOC strength, $I$ is the $2\times 2$ unit matrix, $\sigma_{i}$
is the Pauli matrix, $\Gamma$ is the strength of the Zeeman field, and $c_{
\mathbf{r} \gamma}$ is the atom annihilation operator. $H_{\text{int}} = -g
\int d \mathbf{r} c_{ \mathbf{r} \uparrow }^{\dagger} c_{ \mathbf{r}
\downarrow}^{\dagger} c_{ \mathbf{r} \downarrow } c_{ \mathbf{r} \uparrow } $
is the $s$-wave scattering interaction with $g=4\pi \hbar ^{2}\bar{a}_{s}/m$%
, and the scattering length $\bar{a}_{s}$ can be tuned by the Feshbach
resonance.

The grand partition function at temperature $T$ is $Z = \int D[\psi,\bar{\psi%
} ] \exp\left( -S[\psi,\bar{\psi}] \right)$ with action
\begin{equation}
S[\psi,\bar{\psi}] = \int d\tau d\mathbf{r} \left[ \sum_{\gamma} \bar{\psi}%
_\gamma(\mathbf{r},\tau) \frac{\partial}{\partial \tau} \psi_\gamma(\mathbf{r%
},\tau) + \mathcal{H}(\mathbf{r},\tau) \right].
\end{equation}
Using the standard Hubbard-Stratanovich transformation that introduces the
pairing potential $\Delta(\mathbf{r},\tau) = g \langle \psi_\downarrow(%
\mathbf{r},\tau) \psi_\uparrow(\mathbf{r},\tau) \rangle$ and integrating
over the fermion variables lead to the effective action
\begin{equation}
S_\mathrm{eff} = \int d\tau d\mathbf{r} \left[ \frac{|\Delta(\mathbf{r}%
,\tau)|^2}{g} - \frac{T}{2V} \ln \mathrm{det} \frac{\mathbf{M}}{T} + \xi
\delta(\mathbf{r} - \mathbf{r}^\prime ) \right],
\end{equation}
where $\xi = -\hbar^2\nabla^2 / 2m - \mu$ and the matrix $\mathbf{M}$ is
\begin{equation}
\mathbf{M} =
\begin{pmatrix}
\partial_\tau + \tilde\xi_\uparrow & h_\perp & 0 & -\Delta \\
-h_\perp^\ast & \partial_\tau + \tilde\xi_\downarrow & \Delta & 0 \\
0 & \Delta^\ast & \partial_\tau - \tilde\xi_\uparrow & h_\perp^\ast \\
-\Delta^\ast & 0 & h_\perp & \partial_\tau - \tilde\xi_\downarrow%
\end{pmatrix}%
,
\end{equation}
where $h_\perp = -i\hbar \alpha (\partial_y + i \partial_x)$ corresponds to
the transverse component of the SOC field, $\Gamma$ to the parallel
component with respect to the quantization axix $z$, and $%
\tilde\xi_{\uparrow,\downarrow} = \xi \pm \Gamma $.

To proceed, we use the saddle point approximation $\Delta(\mathbf{r},\tau) =
\Delta_0 + \eta(\mathbf{r},\tau)$, and write $\mathbf{M} = \mathbf{M}_0 +
\mathbf{M}_F$. The matrix $\mathbf{M}_0$ is obtained via the saddle point $%
\Delta(\mathbf{r},\tau) \rightarrow \Delta $ which takes $\mathbf{M}
\rightarrow \mathbf{M}_0$, and the functional matrix $\mathbf{M}_F = \mathbf{%
M} - \mathbf{M}_0$ depends only on $\eta(\mathbf{r},\tau)$ and its Hermition
conjugate. Thus, we write the effective action as $S_\text{eff} = S_0 + S_F$%
. The first term is
\begin{equation}
S_0 = \frac{V}{T} \frac{|\Delta|^2}{g} - \frac{1}{2} \sum_{ \mathbf{k}
,i\omega_n,\eta,\lambda} \ln \left( \frac{i\omega_n - E_{ \mathbf{k}
,\eta}^\lambda}{T} \right) + \sum_ \mathbf{k} \frac{\tilde\xi}{T},
\end{equation}
where $\omega_n = (2n+1) \pi T$. Here $E_{ \mathbf{k} ,\eta}^\lambda$ are
the eigenvalues of
\begin{equation}
\mathbf{M}_ \mathbf{k} =
\begin{pmatrix}
\xi_ \mathbf{k} + \Gamma & - h_\perp( \mathbf{k} ) & 0 & -\Delta \\
- h_\perp^\ast( \mathbf{k} ) & \xi_ \mathbf{k} - \Gamma & \Delta & 0 \\
0 & \Delta^\ast & -\xi_ \mathbf{k} - \Gamma & h_\perp(- \mathbf{k} ) \\
-\Delta^\ast & 0 & h_\perp(- \mathbf{k} ) & - \xi_ \mathbf{k} + \Gamma%
\end{pmatrix}%
,
\end{equation}
which describes the Hamiltonian of elementary excitations in the 4-component
Nambu spinor $\Psi_ \mathbf{k} = (c_{ \mathbf{k} \uparrow},c_{ \mathbf{k}
\downarrow},c_{- \mathbf{k} \uparrow}^{\dagger},c_{- \mathbf{k}
\downarrow}^{\dagger})^{\dagger} $. Here, $h_\perp ( \mathbf{k} ) = \alpha
(k_y + ik_x)$ and the quasiparticle excitation energy
\begin{equation}  \label{dispersion}
E_{ \mathbf{k} \pm}^{\lambda} = \lambda \sqrt{ \xi_{ \mathbf{k} }^{2} +
\alpha^{2}k_{\perp}^{2} + \Gamma^{2} + |\Delta|^{2} \pm 2E_{0} }
\end{equation}
is the eigenvalue of $M_ \mathbf{k} $, where $\lambda =\pm $ correspond to
the particle and hole branches, $E_{0} = \sqrt{ \Gamma^{2} (\xi_{ \mathbf{k}
}^{2}+|\Delta|^{2}) + \alpha^{2}k_{\perp}^{2}\xi_{ \mathbf{k} }^{2} } $, and
$k_{\perp} = \sqrt{k_{x}^{2}+k_{y}^{2}} $. For $\alpha =\Gamma =0$, Eq.~(\ref%
{dispersion}) reduces to $E_{ \mathbf{k} }^{\lambda} = \lambda \sqrt{ \xi _{
\mathbf{k} }^{2}+|\Delta|^{2} } $ in the standard BCS theory.

The saddle-point grand potential function $\Omega = -T \ln Z_0$ is
\begin{equation}
\Omega= V \frac{|\Delta|^2}{g} - \frac{T}{2} \sum_{ \mathbf{k}
,\eta,\lambda} \ln \left( 1 + e^{-E_{ \mathbf{k} ,\eta}^\lambda/T} \right) +
\sum_ \mathbf{k} \xi_ \mathbf{k} ,
\end{equation}
leading to the order parameter via minimization of $\Omega$ with respect to $%
\Delta$:
\begin{equation}  \label{eq-kfas}
\frac{\Delta}{g} + \frac{1}{2V} \sum_{ \mathbf{k} ,\eta,\lambda} f(E_{
\mathbf{k} ,\eta}^\lambda ) \left( \frac{\partial E_{ \mathbf{k}
,\eta}^\lambda}{\partial \Delta^\ast} \right) =0,
\end{equation}
where $f(x) = 1/(1+\exp(-x/T))$ is the Fermi-Dirac distribution function.
For the case of 2D gas, Eq.~(\ref{eq-kfas}) can be expressed as
\begin{equation*}
\sum_{ \mathbf{k} ,\eta} \left[ \left( 1-\eta \Gamma^{2}/E_{0} \right) \tanh
(E_{ \mathbf{k} ,\eta}^{+}/2T)/4E_{ \mathbf{k} ,\eta}^{+} - \frac{1}{%
2\epsilon_{ \mathbf{k} } + E_b} \right] = 0,
\end{equation*}
where $E_b = \hbar^2/m a^2$ is the binding energy controlled by Feshbach
resonances, and for 3D gas,
\begin{equation*}
\frac{V m}{4\pi \hbar^{2}a_{s}} = -\sum_{ \mathbf{k} ,\eta} \left[ \left(
1-\eta \Gamma^{2}/E_{0} \right) \tanh (E_{ \mathbf{k} ,\eta}^{+}/2T)/4E_{
\mathbf{k} ,\eta}^{+} - \frac{1}{2\epsilon_{ \mathbf{k} }} \right].
\end{equation*}
The ultra-violet divergence at the large $ \mathbf{k} $ in Eq.~(\ref{eq-kfas}%
) has been regularized through subtracting the term $1/2\epsilon_{ \mathbf{k}
}$ and $a_{s}$ is defined as the renormalized scattering length. The total
number of atoms can be obtained from $N = - \left( \partial \Omega /
\partial \mu \right)$:
\begin{equation}  \label{eq-n}
N = \sum_{ \mathbf{k} ,\eta} \left[ 1 + \left( \frac{\eta
(\alpha^{2}k_{\perp}^{2}+\Gamma^{2})}{E_{0}} - 1 \right) \frac{\xi_{ \mathbf{%
k} }}{2E_{ \mathbf{k} ,\eta}^{+}} \tanh (E_{ \mathbf{k} ,\eta}^{+}/2T) %
\right].
\end{equation}
We self-consistently solve the order parameter equation (\ref{eq-kfas}) and
the number equation (\ref{eq-n}) for different parameters $\left( \alpha
K_{F},\Gamma,\nu,T\right)$ for a fixed atom density $n$ to determine $\Delta$
and $\mu$. Here $\nu = 1/K_{F}a_{s}$ and $K_{F}=\left( 3\pi
^{2}n\right)^{1/3}$ is the Fermi momentum for a non-interacting 3D Fermi gas
with the same density at $\Gamma = \alpha =0$ and $\nu = E_b$ and $K_F =
(2\pi n)^{1/2}$ for a non-interacting 2D Fermi gas. The energy unit is
chosen as the Fermi energy $E_{F}=\hbar^{2}K_{F}^{2}/2m$.


\section{Topological Phase transitions during BCS-BEC Crossover ($T=0$)}



\subsection{Phase diagram}


\begin{figure}[t]
\includegraphics[width = 1\linewidth]{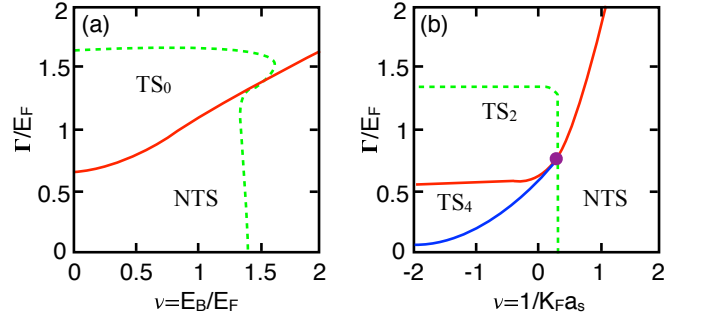}
\caption{ (Color online) Plot of ($\Gamma-\protect\nu$) phase diagram for
(a) 2D and (b) 3D Fermi gas at $T=0$ with $\protect\alpha K_F = 0.4$. Here, $%
\protect\nu$ represents interaction parameters, $E_B/E_F$ for 2D and $1/K_F
a_s$ for 3D systems. We label fully gapped non-topological superfluid phase
as NTS, fully gapped topological superfluid phase as TS$_0$, and topological
superfluids with 2 and 4 nodes as TS$_2$ and TS$_4$, respectively. Red and
blue lines denote $\Gamma_c = \protect\sqrt{\protect\mu^2 + \Delta^2}$ and $%
\Gamma_\Delta = \Delta$, respectively, where the topological phase
transitions occur. Green dashed lines correspond to $\protect\mu = 0$, below
which $\protect\mu >0$ and above which $\protect\mu<0$. Note that TS$_4$
superfluid phase, which is characterized by 4 nodes along $z$-axis, arises
when $\protect\mu > 0$ and merges at the tricritical point, which is
indicated by purple circle in the phase diagram of 3D fermion superfluid. }
\label{fig-1}
\end{figure}
In Fig.~\ref{fig-1}, we show the mean-field zero temperature phase diagram
as a function of the Zeeman field $\Gamma$ and the interaction parameter $\nu
$ in the presence of Rashba SOC ($\alpha K_F = 0.4$). We label the uniform
topologically non-trivial $s$-wave superfluid phases with zero, one, two,
and four nodal points as TS$_0$, TS$_1$, TS$_2$, and TS$_4$, respectively.
For 2D Fermi gas as in Fig.~\ref{fig-1} (a), topologically non-trivial
superfluid phase with one node, TS$_1$, occurs only along $\Gamma = \Gamma_c
= \sqrt{\mu^2 + \Delta^2}$ line. Even though below and above $\Gamma=\Gamma_c
$, the superfluid gap remains finite everywhere in the momentum space, thus
the excitation spectrum is fully gapped, $\Gamma_c$ is a phase boundary
between the non-topological superfluid (NTS) and non-trivial (TS$_0$)
superfluid phases. Even though the TS$_0$ phase is fully gapped, the phase
is topologically non-trivial because in the presence of an order parameter
defect such as a vortex, the vortex core supports a non-degenerate zero
energy Majorana fermion excitation. The TS$_0$ phase has recently received
wide attention in condensed matter physics ~\cite{Sau-2010,Sau-2010-PRB,Roman,Oreg,Mao1,Mao2,Mourik12,Deng12,Das12}  where 2D semiconducting thin
films and 1D semiconducting nanowires with proximity induced $s$-wave
superconductivity and a Zeeman field have been shown to support Majorana
fermions in order parameter defects such as vortices and boundaries. Such
localized Majorana fermion excitations are thought to support non-Abelian
statistics which has been proposed as a key resource for fault-tolerant
topological quantum computation (TQC) \cite{Nayak-2008}. 2D ultra cold fermionic gases with
laser induced Rashba SOC and a Zeeman field in the TS$_0$ phase can also
support Majorana fermions in vortices and sample boundaries and thus can
potentially act as a platform for fault tolerant TQC. In contrast, for 3D
Fermi superfluid as in Fig.~\ref{fig-1} (b), the blue line determined by the
Chandrasekhar-Clogston like condition $\Gamma = \Delta$ is a topological
phase boundary between fully gapped NTS ($\Gamma < \Delta$) and gapless TS ($%
\Gamma > \Delta$) phases. In addition, within the gapless TS phase, $\Gamma_c
$ is a quantum critical point of the topological phase transition between
the TS phases with two (TS$_2$) and four (TS$_4$) nodes. The fermions with
linear dispersion relations near each node in the TS$_2$ and TS$_4$ phases
are a manifestation of the relativistic Weyl fermions. The bulk nodal
excitation spectra of these phases are also topologically protected and a
topological invariant characterizing the nodal points (in the context of
He3) was formulated by Volovik \cite{Volovik-2003}. Recently, similar topologically protected
Weyl fermion modes have been shown to exist also in condensed matter systems
such as ferromagnetic superconductors \cite{Sau-2012}.


\subsection{Chemical potential}


Chemical potential can be calculated by solving the order parameter equation
with fixing the total number of the system self-consistently. With Zeeman
field $\Gamma$ and interaction $\nu$ as tuning parameters, one can have
chemical potential as a function of $\Gamma$ and $\nu$, that is, $\mu =
\mu(\Gamma, \nu)$.

\begin{figure}[t]
\includegraphics[width=1.0\linewidth]{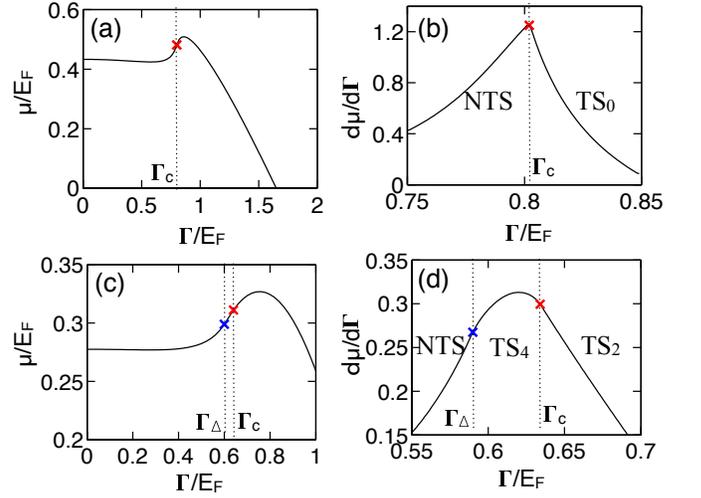}
\caption{ (Color online) Plot of chemical potential $\protect\mu$ and its
derivative with respect to $\Gamma$, $\frac{d\protect\mu}{d\Gamma}$ for
(a)-(b) 2D Fermi gas at $E_b = 0.5 E_F$ and (c)-(d) 3D Fermi gas at $1/K_F
a_s = 0$ with $\protect\alpha K_F = 0.4$. Blue and red crosses indicate the
critical points $\Gamma_\Delta = \Delta$ and $\Gamma_c = \protect\sqrt{%
\protect\mu^2+\Delta^2}$, respectively. One finds that fully gapped NTS
phase occupies at $\Gamma < \Gamma_c$ for 2D and $\Gamma <\Gamma_\Delta$ for
3D gases, respectively. }
\label{fig-2}
\end{figure}

In Fig.~\ref{fig-2}, we show the chemical potential as a function of $\Gamma$ for a
given interaction $\nu$ and SOC $\alpha K_F$. Fig.~\ref{fig-2} (a) and (b) are plots
of the chemical potential $\mu$ and its derivative with respect to $\Gamma$ $%
,\frac{d \mu}{d\Gamma}$, for two dimensional Fermi gas at $E_b = 0.5 E_F$
and $\alpha K_F = 0.4$. In the side of low Zeeman field $\Gamma$, where $%
\Gamma < \Gamma_c = \sqrt{\mu^2 + \Delta^2}$ and the system is in the
non-topological superfluid phase (NTS), $\mu$ is slowly varying as $\Gamma$
increases. In the region of strong Zeeman field, where $\Gamma > \Gamma_c$
and the system is in a gapped TS$_0$ phase, $\mu$ decreases fast with
increasing $\Gamma$. Near the quantum critical point $\Gamma_c$, the
slope changes discontinuously as can be seen clearly in the plot of $\frac{%
d\mu}{d\Gamma}$ in Fig.~\ref{fig-2} (b), indicating the change in the curvature.

As for the 3D Fermi superfluids, topological phase transition from a fully
gapped NTS to a gapless TS$_4$ state occurs at Zeeman field $\Gamma =
\Gamma_\Delta = \Delta$, and TS$_4$ to TS$_2$ phase transition occurs at $%
\Gamma = \Gamma_{c} = \sqrt{\mu^2 + \Delta^2}$. Fig.~\ref{fig-2} (c) and (d)
illustrate $\mu$ as a function of $\Gamma$ at unitarity ($1/K_F a_s = 0$)
with the same SOC strength as that in 2D fermion gas. Even though the
signatures of topological phase transitions across the phase boundaries are
weaker than the one for the 2D Fermi system, $\mu$ in the fully gapped NTS
phase ($\Gamma < \Gamma_{\Delta}$) increases with increasing $\Gamma$, while
$\mu$ decreases with increasing $\Gamma$ when it is in the TS$_2$ phase ($%
\Gamma > \Gamma_{c}$). The curvature of the chemical potential versus the
Zeeman field changes as the system moves from the gapped to the gapless
phases and vice versa. In contrast to the 2D Fermi superfluid, however, $%
\frac{d\mu}{d\Gamma}$ plot doesn't exhibit a cusp at the phase boundary
because of the existence of the TS$_4$ phase surrounded by the gapped NTS
and gapless TS$_2$ phases. It is when $\mu = 0$ that the two distinct
critical points $\Gamma_{\Delta}$ and $\Gamma_{c}$ merge ($%
\Gamma_{\Delta}=\Gamma_c$) and a tricritical point occurs, leading to the
vanishing of the TS$_4$ phase and the emergence of the cusp in $\frac{d\mu}{%
d\Gamma}$ even in a 3D Fermi superfluid system.

\subsection{Compressibility}

The compressibility can be obtained from the knowledge of the pressure, as
defined by, $\kappa_T = - (1/V) (\partial p / \partial V)$. But in the
grand-canonical ensemble, where we need to fix the average number of
particles, the isothermal compressibility can be directly written as,
\begin{equation}
\kappa_T = \frac{1}{N^2} \left( \frac{\partial N}{\partial \mu}
\right)_{T,V}.
\end{equation}
In the superfluid state, we need to consider $\mu$ dependence of $\Delta$.
From the order parameter equation $\left( \partial \Omega / \partial \Delta
\right)= 0$, we have $(\partial / \partial \mu )(\partial \Omega /\partial
\Delta ) = 0$, leading to
\begin{equation}
\left( \frac{\partial \Delta}{\partial \mu} \right)_{T,V} = -\left( \frac{%
\partial N}{\partial \Delta} \right)_{T,V,\mu} \left( \frac{\partial^2 \Omega%
}{\partial \Delta^2} \right)^{-1}_{T,V,\mu}
\end{equation}
Thus the isothermal compressibility in the superfluid phase can be written
as
\begin{equation}
\kappa_T = \frac{1}{N^2} \left[ \left( \frac{\partial N}{\partial \mu}
\right)_{T,V,\Delta} - \left( \frac{\partial N}{\partial \Delta}
\right)^2_{T,V,\mu} \left( \frac{\partial^2 \Omega}{\partial \Delta^2}
\right)^{-1}_{T,V,\mu} \right]
\end{equation}

\begin{figure}[t]
\includegraphics[width=1.0\linewidth]{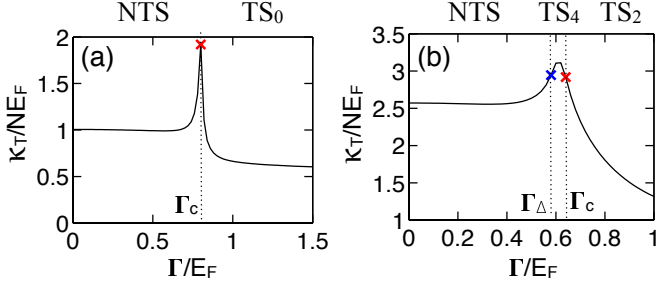}
\caption{ (Color online) Plot of isothermal compressibility $\protect\kappa_T
$ as a function of $\Gamma$ for (a) 2D Fermi gas at $E_b = 0.5 E_F$ and (b)
3D Fermi gas at $1/K_F a_s = 0$ with $\protect\alpha K_F = 0.4$. Blue and
red crosses indicate the critical points $\Gamma_\Delta = \Delta$ and $%
\Gamma_c = \protect\sqrt{\protect\mu^2+\Delta^2}$, respectively, same as in
Fig.~\ref{fig-1} and Fig.~\ref{fig-2}. }
\label{fig-3}
\end{figure}

In Fig.~\ref{fig-3}, we present the isothermal compressibility $\kappa_T$ as a
function of $\Gamma$ for a given interaction $\nu$ at $T=0$ with $\alpha K_F
= 0.4$ to investigate the signature of the topological phase transitions.
Since $\kappa_T$ is related to the first derivative of the total
number $N$ with respect to $\mu$, that is, $\kappa_T \sim \frac{\partial N }{%
\partial \mu}|_{T,V}$, it is expected to show the signatures of the
topological phase transitions as seen in the plots of $\mu$ in Fig.~\ref{fig-2}.
 In Fig.~\ref{fig-3}, $\kappa_T$ shows a decreasing trend in the
superfluid phase with increasing $\Gamma$, while it is an increasing
function of $\Gamma$ in normal phase as will be shown in Fig.~\ref{fig-5}. For 2D
uniform Fermi superfluids, at the critical Zeeman field $\Gamma_c$, a sharp
peak emerges as shown in Fig.~\ref{fig-3} (a). This peak in the compressibility
corresponds to the phase boundary between the 2D gapped NTS and the 2D
gapped TS$_0$ phases. In our calculations we find that the compressibility
provides the strongest signature of the phase transition to the 2D
topological phase with the Majorana fermion excitations. In 2D at zero
temperature there is no additional phase transition from the TS$_0$ phase to
the normal phase at higher values of $\Gamma$. This is because, with a
finite Rashba SOC, the superfluid pair potential always remains non-zero
even at large values of the Zeeman field. When the temperature is included
as another parameter on the phase diagram, a normal phase with $\Delta=0$
emerges even in 2D (see below).

As for the 3D case, Fig.~\ref{fig-3} (b) shows that $\kappa_T(\Gamma)$ again decreases
with increasing $\Gamma$ in the superfluid phase, and shows the signature of
the topological phase transitions by a smoothly peaked curve near $\Gamma =
\Gamma_c$. As already seen in the plot of $\mu$ vs. $\Gamma$, $\kappa_T$ of
3D fermion system in the TS$_4$ phase, which is sandwiched between the
gapped NTS and gapless TS$_2$ phases, is a fast varying function of $\Gamma$%
, while the compressibility in the gapped NTS phase is slowly varying.

\section{Thermodynamic Sigantures ($T \neq 0$)}

Experimentally, the thermodynamic properties of a given substance is
determined by measuring an equation of state (EoS), such as the pressure $%
p(\mu,T)$ as a function of the chemical potential $\mu$ and the temperature $%
T$. Equivalently, replacing the pressure by the density $n = \left(\frac{%
\partial p}{\partial \mu} \right)_T$, one can determine the density EoS, $%
n(\mu,T )$. The local gas density $n(V )$ can be measured \cite{Ku-2012} as a
function of the local trapping potential $V$ from in situ absorption images of a
trapped strongly interacting Fermi gas at a Feshbach resonance. From the
definition of the compressibility $\kappa_T = \frac{1}{N^2} \left( \frac{%
\partial N}{\partial \mu} \right)_T $, the chemical potential as a function
of temperature $T$ is attainable experimentally without numerical
derivatives of the data \cite{Ku-2012}. Then, from
the energy, pressure, and the chemical potential, the entropy $S = \frac{1}{T%
} (E + pV - \mu N)$ and the specific heat $c_V$ can be obtained as a
function of temperature $T$.


\subsection{Thermodynamic Properties}


From the mean-field grand partition function
\begin{equation}
Z = \text{Tr} \left( e^{-(H-\mu N)/T} \right)
\end{equation}
we can calculate the corresponding grand potential function $\Omega = -T \ln
Z = \langle H \rangle - T S - \mu \langle N \rangle = -pV$, where $S$ is the
entropy and $p$ is the pressure of the system. The differential of the grand
potential function is
\begin{equation}
d \Omega = -S dT - p dV - \langle N \rangle d \mu,
\end{equation}
allowing us to obtain interesting thermodynamic properties. The explicit
expression of $\Omega$ in terms of the Hamiltonian under consideration is
\begin{equation}
\Omega = -\frac{T}{2} \sum_{ \mathbf{k} ,\eta,\lambda} \ln \left( 1 +
e^{-E_{ \mathbf{k} ,\eta}^\lambda/T} \right) + \sum_ \mathbf{k} \xi_ \mathbf{%
k} + V\frac{|\Delta|^2}{g}
\end{equation}

Then, the pressure of the ultracold Fermi gas can be obtained through the
grand potential function via $p = -\Omega / V. $ In the thermodynamic limit,
it reads
\begin{equation}
p = -\frac{T}{2} \sum_{\eta,\lambda} \int \frac{d^Dk}{(2\pi)^D} \ln \left( 1
+ e^{-E_{ \mathbf{k} ,\eta}^\lambda/T} \right) + \int \frac{d^Dk}{(2\pi)^D}
\xi_ \mathbf{k} + \frac{|\Delta|^2}{g}
\end{equation}

The entropy can be obtained by
\begin{equation}
S = -\left( \frac{\partial \Omega }{\partial T} \right)_{\mu,V}.
\end{equation}

Then, the heat capacity can be obtained from the knowledge of the entropy.
\begin{equation}
c_{V} = \left( \frac{\partial S}{\partial T} \right)_{\mu,V}.
\end{equation}
In the superfluid phase, $\Delta$ depends on temperature $T$ as
\begin{equation}
\left( \frac{\partial \Delta}{\partial T} \right)_{\mu,V} = - \left( \frac{%
\partial S}{\partial \Delta} \right)_{T,V,\mu} \left( \frac{\partial^2 \Omega%
}{\partial \Delta^2} \right)^{-1}_{T,V,\mu},
\end{equation}
leading to the specific heat in the superfluid phase
\begin{equation}
c_V = \left( \frac{\partial S}{\partial T} \right)_{\mu,V,\Delta} - \left(
\frac{\partial S}{\partial \Delta} \right)^2_{T,V,\mu} \left( \frac{%
\partial^2 \Omega}{\partial \Delta^2} \right)^{-1}_{T,V,\mu}.
\end{equation}


\subsection{Finite Temperature Phase Diagram}

\begin{figure}[t]
\includegraphics[width=1.0\linewidth]{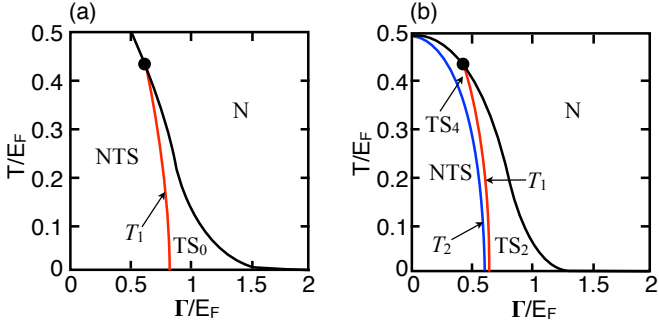}
\caption{ (Color online) Plot of ($\Gamma-T$) phase diagrams for (a) 2D
Fermi gas at $E_b = 0.5E_F$ and (b) 3D Fermi gas \textbf{\ at $1/K_F a_s=0$
with $\protect\alpha K_F = 0.4$.} Red and blue lines denote $\Gamma_c =
\protect\sqrt{\protect\mu^2 + \Delta^2}$ and $\Gamma_\Delta = \Delta$,
respectively. Black lines represent $\Gamma_0$ above which the order
parameter vanishes, or $\Delta = 0$. The tricritical temperatures, at which $%
\Gamma_c$ meets with $\Gamma_0$ (black circles), for 2D and 3D gases are $T
= 0.43E_F$ at $\Gamma_c = \Gamma_0= 0.62E_F$ and $T = 0.44E_F$ at $\Gamma_c
= \Gamma_0= 0.42E_F$,respectively. }
\label{fig-4}
\end{figure}

Fig.~\ref{fig-4} presents mean-field $(T-\Gamma)$ phase diagram for 2D and 3D
homogeneous Fermi gas in the presence of Rashba-type of spin-orbit coupling.
We choose the interaction parameter $\nu$ as a two-body binding energy $E_b
= 0.5E_F$ for 2D Fermi gas (a), while $1/K_F a_s = 0$ for 3D Fermi
gas (b) with Rashba SOC $\alpha K_F = 0.4$. With increasing temperature $T$%
, the superfluid order parameter $\Delta$ decreases and undergoes the second
order phase transition to normal phase ($\Delta =0$) denoted as N at $T=T_c$.

In Fig.~\ref{fig-4} (a), within the superfluid phase of 2D fermion gas, topological
phase transition between the gapped NTS phase ($\Gamma < \Gamma_c$) and the gapped TS$_0$ phase (%
$\Gamma > \Gamma_c$) occurs at $T=T_{1} < T_c$ indicated with a red line. In
addition, the TS$_0$ phase vanishes at the tricritical temperature $T_{1} =
T_c = 0.43 E_F$ with our choice of parameters $E_b$ and $\alpha$. In
Fig.~\ref{fig-4} (b), $T_{2}$ (blue line) determined by the
Chandrasekhar-Clogston-like condition, $\Gamma = \Delta$, is a phase
boundary between the NTS phase ($T < T_{2}$) and the TS$_4$ phase ($T_{2} < T < \text{min}%
[T_c, T_{1}]$). It is when $\Gamma = 0$ that $T_{2} = T_c$ occurs and the system above $T_c$
becomes the normal state ($\Delta = 0$). Thus, for $\Gamma=0$, there is only one finite temperature phase transition at $T_c (=T_2)$ which is a 
 transition between the non-topological superfluid (NTS) ($T < T_c$) and the normal state ($T > T_c$). In a similar manner as in the 2D system, TS$%
_4$ phase vanishes at finite $\Gamma$, where $T_c = T_{1} = 0.43 E_F$ at
unitarity ($1/K_F a_s = 0$).

\begin{figure}[t]
\includegraphics[width=1.0\linewidth]{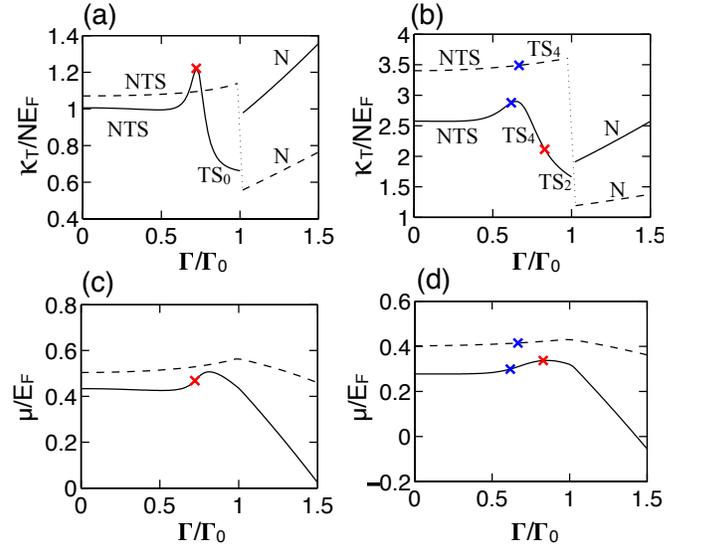}
\caption{ (Color online) Plot of isothermal compressibility $\protect\kappa_T
$ and chemical potential $\protect\mu$ as a function of $\Gamma / \Gamma_c$
for (a),(c) 2D and (b),(d) 3D Fermi gas at temperature $T=0.1 E_F$ (solid)
and $T = 0.45E_F$ (dashed). We indicates topological phase transitions with
red (TS$_4$ to TS$_2$) and blue (NTS to TS$_4$) crosses within the uniform
superfluid phases. One finds that normal (N) to superfluid phase transitions
occur at $\Gamma_0$ and are characterized by the discontinuity of the
isothermal compressibility $\protect\kappa_T$, while the chemical potential $%
\protect\mu$ remains a continuous function of $\Gamma$ as well as a function
of $T$. }
\label{fig-5}
\end{figure}


\subsection{ Thermodynamic Quantities as a function of temperature}

Since we learned that the chemical potential and the isothermal
compressibility at zero temperature can show the evidence of the topological
phase transitions as a function of the tuning parameters such as the Zeeman
field $\Gamma$ and the interaction parameter $\nu$, we first consider below
the chemical potential and the compressibility at finite temperatures.

In Fig.~\ref{fig-5}, we show the chemical potential and isothermal compressibility as
a function of the reduced Zeeman field $\Gamma / \Gamma_0$ for 2D on the
left column and 3D on the right column at both $T> T_{1}$ (dashed) and $T<
T_{1}$ (solid). We choose the Rashba SOC $\alpha K_F = 0.4$ and the
interaction parameters $E_b = 0.5E_F$ for 2D and $1/K_F a_s = 0$ for
3D Fermi gases, respectively. Fig.~\ref{fig-5} (a) and (b) show that, at low
temperatures, $T<T_{1}$, $\kappa_T$ for both 2D and 3D fermions exhibit
peak at or near the topological critical point, and a discontinuity at the
phase transition to the normal state at $\Gamma = \Gamma_0$. Meanwhile, at
high temperatures (dashed), $T > T_{1}$, it is hard to see the signature of
the topological phase transition between the gapped NTS and gapless TS$_4$
phases. We notice that as temperature increases $\kappa_T$ increases in the
superfluid phase, while it decreases in the normal phase. Fig.~\ref{fig-5} (c) and (d)
are plots of both 2D and 3D chemical potentials as a function of $\Gamma
/\Gamma_0$ for $T=0.1E_F$ (solid) and $T=0.45E_F$ (dashed), respectively. It
shows that $\mu$ is an increasing function of $T$ and $\mu$ in the
superfluid phase is varying slowly compared with that in normal phase. Even
though the signatures of the topological transition, which are strong at zero
temperature, are weakened by the thermal fluctuations, the characteristic
features for the variations of the thermodynamic quantities near the phase transitions remain the same.

\begin{figure}[t]
\includegraphics[width=1.0\linewidth]{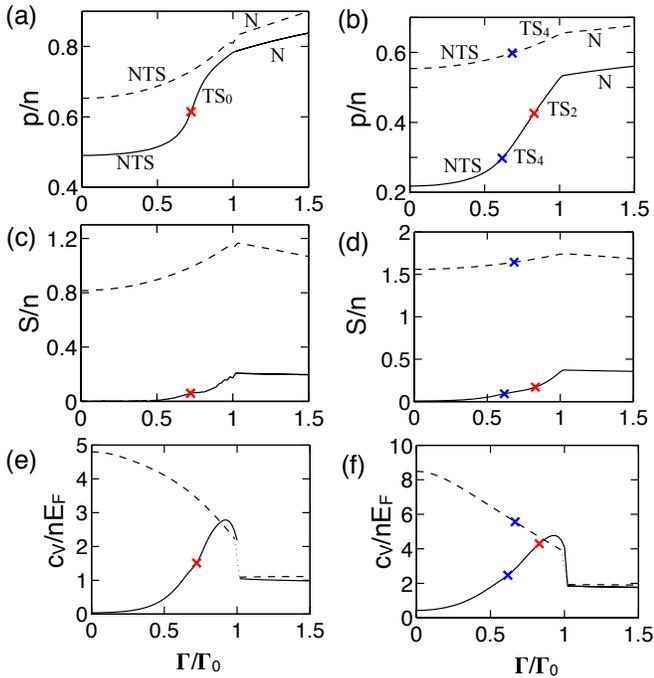}
\caption{ (Color online) Plot of pressure $p$, entropy $S$, and specific
heat $c_V$ as a function of $\Gamma / \Gamma_0$, where $\Gamma_0$
corresponds to the Zeeman field at $\Delta =0$, for (a),(c),(d) 2D Fermi gas
at $E_b = 0.5 E_F$ and (b),(d),(f) 3D Fermi gas at $1/K_F a_s = 0$ at $T =
0.1E_F$ (solid) and $T=0.45E_F$ (dashed). We used the red and blue crosses
for $\Gamma_c$ and $\Gamma_\Delta$, respectively. One notices that the
signatures of the normal (N) to superfluid phase transitions are visible in
both 2D and 3D Fermi gases, characterized by the suppression of $p$ and $S$
in the superfluid state and the discontinuity of $c_V$. On the other hand,
it is difficult to observe the topological phase transitions even with some
weak signatures. }
\label{fig-6}
\end{figure}

In Fig.~\ref{fig-6}, we present the other relevant thermodynamic quantities in 2D and
3D Fermi gas, such as pressure $p$, entropy $S$, and specific heat $c_V$,
which can be observed experimentally, as a function of $\Gamma / \Gamma_0$
at $T = 0.1 E_F$ (solid) and $T=0.45 E_F$ (dashed). We choose $E_b = 0.5E_F$ for
2D and $1/K_F a_s = 0$ for 3D Fermi system with Rashba SOC $\alpha K_F = 0.4$. 
The main conclusion from these calculations is that the signatures of the
topological phase transitions are much weaker in these quantities than those
in the case of ordinary phase transitions from the ordinary $s$-wave
superfluid to normal phases.

\section{Conclusions}

In summary, we have investigated the two- and three-dimensional uniform
fermion $s$-wave superfluids in the presence of Rashba spin-orbit coupling
together with Zeeman field as they undergo topological phase transitions as
a function of the interaction parameter $\nu$ and Zeeman field $\Gamma$ at
zero temperature and as a function of Zeeman field $\Gamma$ at finite
temperatures. In 2D $s$-wave Fermi gas, for $\Gamma = \Gamma_c = \sqrt{\mu^2
+ \Delta^2}$ the superfluid phase is identified with a gapless topological
superfluid (with one node at $k=0$), while it becomes a fully gapped
non-topological superfluid phase at $\Gamma < \Gamma_c$ and a fully gapped topological
superfluid phase TS$_0$, capable of supporting Majorana fermion excitations at
order parameter defects such as vortices, at $\Gamma > \Gamma_c$. The TS$_0$ phase has recently received
wide attention in condensed matter physics ~\cite{Sau-2010,Sau-2010-PRB,Roman,Oreg,Mao1,Mao2,Mourik12,Deng12,Das12}  where 2D semiconducting thin
films and 1D semiconducting nanowires with proximity induced $s$-wave
superconductivity and a Zeeman field have been shown to support Majorana
fermions in order parameter defects such as vortices and boundaries. Such
localized Majorana fermion excitations are thought to support non-Abelian
statistics which has been proposed as a key resource for fault-tolerant
topological quantum computation (TQC) \cite{Nayak-2008}. In
contrast, 3D $s$-wave Fermi gas is a fully gapped non-topological superfluid
at $\Gamma < \Delta$, becomes a gapless topological superfluid with four
topologically protected Weyl fermion points \cite{Volovik-2003} at $\Delta < \Gamma < \Gamma_c$
and another topological superfluid with two topologically protected Weyl fermion points at $\Gamma >
\Gamma_c$. With our calculated mean-field phase diagrams in both 2D and 3D we investigated the
thermodynamic properties across the relevant topological phase transitions,
including the isothermal compressibility, the chemical potential, the
pressure, the entropy, and the specific heat as a function of the Zeeman
field $\Gamma$ and the interaction parameter $\nu$. In particular, the
isothermal compressibility of the system supporting the Majorana fermions,
which is the 2D fermion gas with Rashba SOC together and a perpendicular
Zeeman field, reveals the topological phase transition via a strong peak at
the critical point $\Gamma = \Gamma_c$ (Fig.~\ref{fig-3}). Furthermore, the phase boundary can also be
identified by the chemical potential through the change of the curvature as
a function of $\Gamma$ (Fig.~\ref{fig-2}). For the 3D Fermi gas, the topological phase
transition from the fully gapped non-topological phase to the gapless
topological phases can be probed by the position of the peak in the
isothermal compressibility as a function of $\Gamma$, even if the signatures
of the topological transitions are weaker than those in the 2D system. The
signatures of the topological phase transitions at low temperature
disappear as the temperature goes above the critical temperature $T_{1}$. The
other thermodynamic quantities, such as the pressure, the entropy, and the
specific heat, have only weak signatures of the topological phase
transitions (that is, the transition between a conventional (non-topological) superfluid and a topological superfluid such as the Majorana and the Weyl phases), while these are useful quantities to identify the
phase transitions from the normal phase to the topological superfluid phases with decreasing values of the Zeeman field. 

\acknowledgements{This work is supported by AFOSR (FA9550-13-1-0045), DARPA MTO (FA9550-10-1-0497), NSF-PHY (1104546), and ARO (W911NF-12-1-0334).
and


\end{document}